\documentclass[11pt,twoside]{article}
\usepackage{./asp2014}
\usepackage{color}

\hypersetup{
 colorlinks= true, 
 urlcolor  = black, 
 linkcolor = red, 
 citecolor = blue 
} 

\aspSuppressVolSlug
\resetcounters

\begin{document}

\title{Accreting Supermassive Black Holes in Nearby Low-mass Galaxies}

\author{Kristina Nyland$^1$ and Katherine Alatalo$^2$ 
\\
\affil{$^1$National Radio Astronomy Observatory, Charlottesville, VA, 22903; \email{knyland@nrao.edu}}
\affil{$^2$Space Telescope Science Institute, 3700 San Martin Dr., Baltimore, MD 21218, USA; \email{kalatalo@stsci.edu}}}


\begin{abstract}
The ngVLA will facilitate deep surveys capable of detecting the faint and compact signatures of accreting supermassive black holes (SMBHs) with masses below one million solar-masses hosted by low-mass ($< 10^9$ solar-masses) galaxies.  This will provide important new insights on both the origins of supermassive black holes and the possible impact of active galactic nucleus-driven feedback in a currently unexplored mass regime. 
\end{abstract}

\section{Introduction and Motivation}
It is now well-established that supermassive black holes (SMBHs) with masses ranging from $10^6 - 10^{10} \, M_{\odot}$ commonly reside in the nuclei of galaxies.  A number of lines of evidence, including the observed scaling relations between SMBHs and their hosts, further suggest that the formation and growth of SMBHs and their hosts are inextricably linked \citep{kormendy+13, heckman+14}.  Despite their prevalence and importance to our understanding of galaxy evolution, the origin of supermassive SMBHs at high redshift (e.g., hierarchical merging vs.\ direct gas collapse) and their formation efficiency remain an open areas of research \citep{volonteri+10, bellovary+11, shirakata+16,latif+16}.  Thus, a key observational parameter for constraining the formation of SMBHs is the mass distribution of so-called SMBH ``seeds''  \citep{greene+12, reines+15, mezcua+16}. 

\subsection{Insights on SMBH Formation from Nearby Low-mass Galaxies}

Knowledge of the shape of the SMBH seed mass distribution at high redshift ($z > 6$) would provide strong constraints on the dominant SMBH formation channel, thus elucidating a key missing element in our understanding of SMBH-galaxy co-evolution.  While direct mass measurements of SMBH seeds in the early universe are not yet feasible \citep{reines+16b, volonteri+17}, the detection of their radiative signatures provides information on the SMBH occupation fraction as well as indirect SMBH mass constraints.   

Thus far, studies of high-$z$ SMBHs have been limited to luminous quasars, which may have formed via atypical processes to build the massive SMBHs needed to power their central engines at such early cosmic epochs (e.g., \citealt{mortlock+11}).  The population of accreting SMBHs with $M_{\mathrm{BH}} < 10^6$ M$_{\odot}$ in the local Universe residing in lower-mass and dwarf galaxies offers an alternative means for studying SMBH seed formation 
since these systems have not experienced substantial merger or accretion driven growth (e.g., \citealt{mezcua+17}).  

\subsection{SMBH-driven Feedback in Low-mass Galaxies?}
In addition to their importance for constraining SMBH seed formation and growth scenarios, accreting SMBHs hosted by low-mass galaxies may also impact their hosts though energetic feedback.  While supernova-driven feedback is believed to be the dominant regulatory mechanism associated with low-mass galaxies (e.g., \citealt{martin-navarro+18}), recent observational and theoretical evidence suggests that active galactic nuclei (AGNs) powered by the sub-million-solar-mass SMBHs residing in their nuclei may also be capable of producing significant energetic feedback.  
From a theoretical standpoint, analytical models suggest that AGN feedback in low-mass galaxies may provide an efficient mechanism for the displacement of gas from the host galaxy \citep{silk+17, dashyan+18}, though this possibility remains controversial (e.g., \citealt{trebitsch+18}). This scenario may be most plausible in galaxies with stellar masses in the range of $10^7 \lesssim  M_{*} \lesssim 10^9$~M$_{\odot}$ when preceded by substantial supernova feedback capable of rarefying the ISM, thus making it more susceptible to disruption via AGN feedback \citep{prieto+17, hartwig+18}.    



Observational evidence for AGNs hosted by low-mass galaxies has become increasingly common over the past decade \citep{barth+08, greene+07b, reines+13, moran+14, lemons+15, sartori+15, nucita+17, mezcua+18}, 
though the identification of AGN-driven feedback in low-mass and dwarf galaxies has remained challenging due to the resolution and sensitivity limitations of existing telescopes.  However, recent spatially-resolved optical emission line studies have provided tentative support for AGN feedback in the low-mass-galaxy regime (e.g., \citealt{penny+18}).  

At radio frequencies, the identification of dwarf galaxies harboring jetted AGNs with the potential to impart feedback on their hosts is also challenging with current instruments such as the Karl G. Jansky Very Large Array (JVLA; \citealt{nyland+16, padovani+16}), which lack adequate collecting area and angular resolution to detect their faint, compact signatures.  Despite their importance for probing the impact of jet-driven feedback physics in the low-mass regime, only a handful of candidate jetted AGNs hosted by low-mass galaxies with $M_{\rm BH} < 10^6 \,M_{\odot}$ are known, including NGC\,4395 \citep{wrobel+06}, Henize 2-10 \citep{reines+11, reines+13}, and NGC\,404 \citep{nyland+17}.  

\begin{figure*}[t!]
\centering
\includegraphics[clip=true, trim=0cm 5cm 0cm 5cm, width=\textwidth]{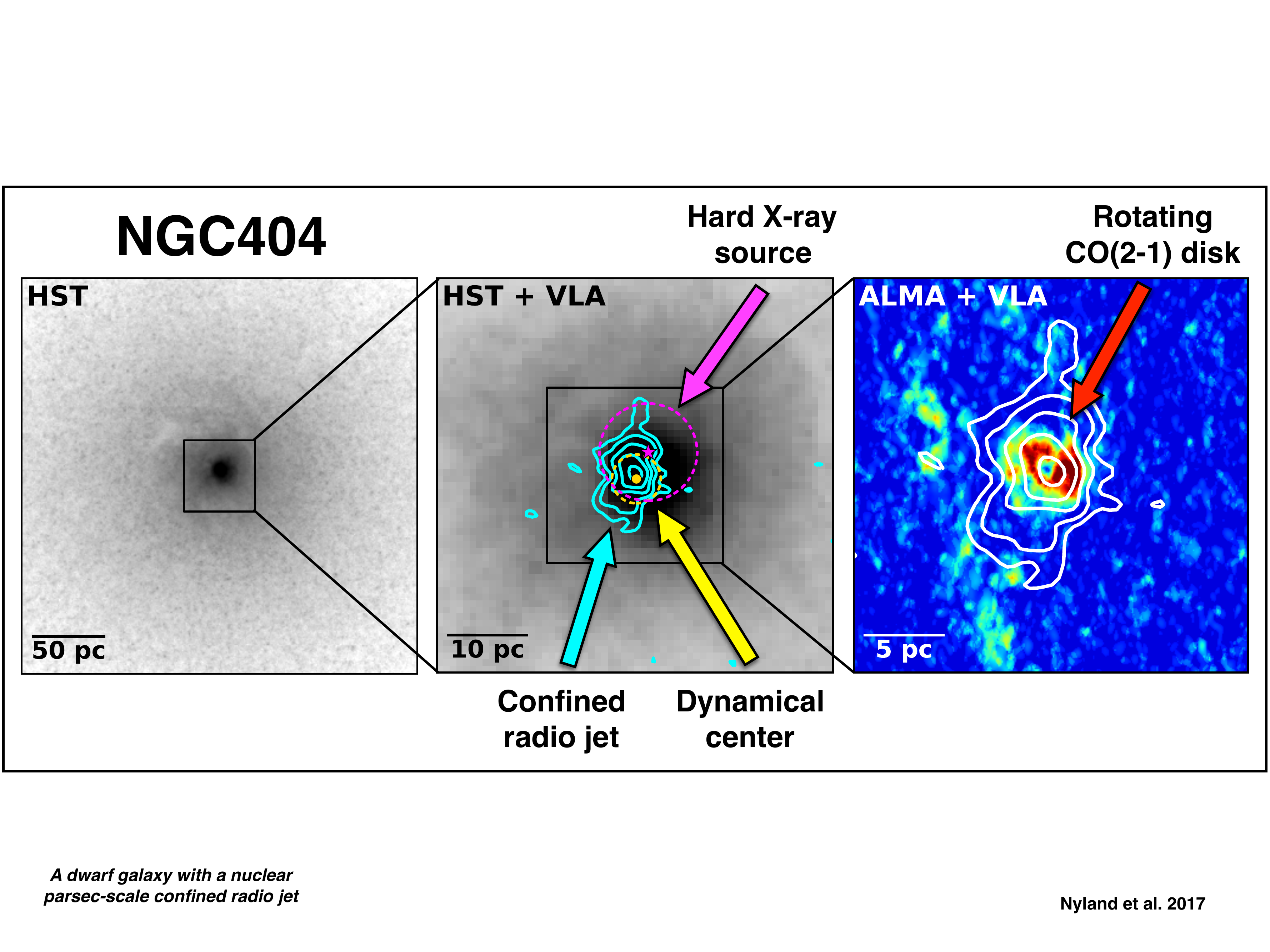} 
\caption{{\it HST}, VLA, {\it Chandra} and ALMA data identify a confined radio jet possibly interacting with the ambient ISM in the nucleus of the dwarf galaxy NGC\,404.  Adapted from \citet{nyland+17}.
}
\label{fig:radio_alma_hst}
\end{figure*}

\subsection{NGC\,404: A Case Study}
In Figure~\ref{fig:radio_alma_hst}, we illustrate the multiwavelength nuclear properties of the nearby dwarf galaxy NGC\,404, which harbors a nuclear radio jet with an extent of $\sim$10~pc characterized by a steep radio spectral index from 1 to 18~GHz ($\alpha \sim -1$; \citealt{nyland+17}).  The spatial coincidence of the radio source, hard X-ray source, dynamical center of NGC\,404, and the rotating circumnuclear CO(2--1) disk are most consistent with a confined radio jet launched by the central low-mass ($M_{\rm BH}<10^5 \, M_{\odot}$ from stellar dynamical modeling; \citealt{nguyen+17}), low-Eddington-ratio ($L_{\rm bol}/L_{\rm Edd} \sim 10^{-6}$; \citealt{nyland+12}) SMBH that has been disrupted by the ambient interstellar medium (ISM).  Multiwavelength evidence for shock excitation supports the possibility of an interaction between the confined jet and the ISM.  This includes the detection of extended H$\alpha$ and soft X-ray emission from {\it HST} and {\it Chandra}, respectively, as well as strong [Fe II] at $26\,\mu$m and rotational H$_2$ emission in archival {\it Spitzer} spectroscopic data \citep{nyland+17}.  

The nucleus of NGC\,404 therefore offers a unique local laboratory for directly constraining the energetic impact on the ambient ISM by a jetted AGN hosted by a dwarf galaxy.  However, the identification of additional jetted AGNs in low-mass galaxies will ultimately be needed to place NGC\,404 in the broader context of galaxy evolution.  Given the limited sensitivity and resolution of current radio telescopes, building a larger sample of jetted low-mass AGNs must await the availability of next generation instruments such as the Next-generation Very Large Array (ngVLA).
\section{Prospects for the ngVLA}
\subsection{Demographics and Energetics of a Hidden SMBH Population}
The ngVLA, with its roughly order of magnitude increase in sensitivity and angular resolution compared to the JVLA, will greatly improve our ability to study compact radio sources in the nuclei of low-mass galaxies associated with SMBH accretion.  The local volume of nearby galaxies within $\sim$10~Mpc encompasses 869 galaxies, about 75\% of which are dwarf galaxies \citep{karachentsev+13}.  Recent studies (e.g., \citealt{mezcua+16}) have suggested that accreting SMBHs with masses in the range of $10^3 \lesssim  M_{\mathrm{BH}} \lesssim 10^6$~M$_{\odot}$ may commonly reside in the nuclei of nearby low-mass galaxies, thus motivating deep searches for the radio continuum signatures of this population.  However, identifying SMBHs in this population of galaxies is inherently difficult due to their expected low masses (M$_{\rm BH} < 10^6 \, M_{\odot}$) and faint accretion signatures.  

The identification of NGC\,404 analogs in other dwarf or low-mass galaxies would offer new insights into the occupation fraction of SMBHs analogous to the SMBH seeds that formed at high redshift, thus profoundly impacting our understanding of the origin of SMBHs.  Deep, high-angular-resolution observations with the ngVLA will both help constrain the SMBH seed mass distribution\footnote{Additional considerations on the topic of constraining SMBH seed formation with the ngVLA are provided in Plotkin \& Reines in this volume.} and also provide new constraints on the energetic impact of AGN feedback associated with SMBHs with masses below one million solar-masses.  These new insights will provide crucial tests of our understanding of SMBH-galaxy co-evolution.  

\begin{figure*}[t!]
\centering
\includegraphics[clip=true, trim=0.05cm 7cm 0.05cm 4cm, width=\textwidth]{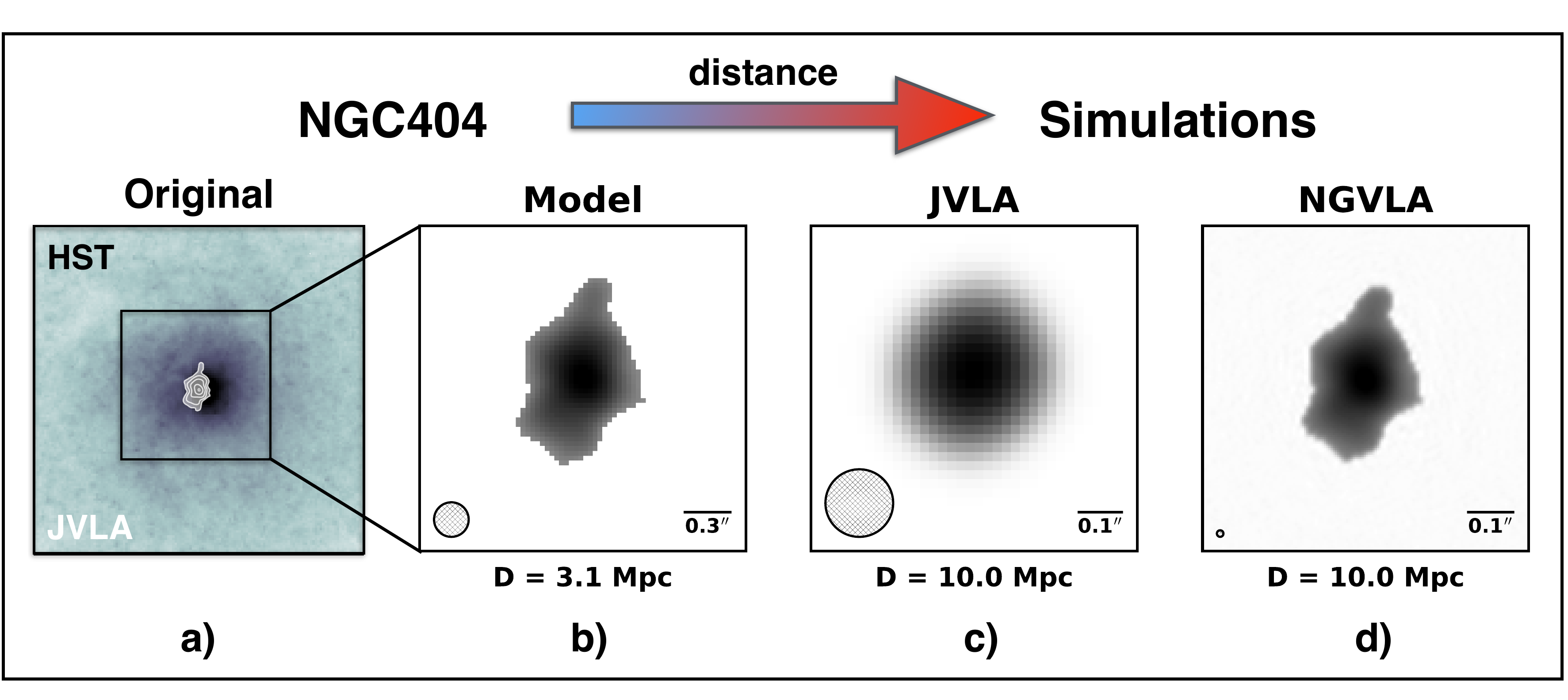}
\caption{Simulated VLA and ngVLA images of an analog to the jetted AGN hosted by the nearby dwarf galaxy NGC\,404 as it would appear if it were $\sim$3X more distant.  {\bf a)} An {\it HST} optical image is shown in the background colorscale and the VLA $Ku$-band (15~GHz) radio contours are overlaid in white.  The angular resolution of the radio data is $\theta_{\mathrm{FWHM}} = 0.20^{\prime \prime}$ and the extent of the radio jets from end to end is 1.13$^{\prime \prime}$ (17~pc).  The NGC\,404 VLA data were originally published in \citet{nyland+17}.  {\bf b)} Model radio image of NGC\,404 based on the original data shown in panel a) and shown at the true source distance of $D = 3.1$~Mpc. {\bf c)}  Simulated VLA A-configuration map of an NGC\,404 analog shifted to a distance of $D = 10.0$~Mpc at 15~GHz with $\theta_{\mathrm{FWHM}} = 0.15^{\prime \prime}$.  {\bf d)} Simulated map of an NGC\,404 analog at $D = 10.0$~Mpc as it would appear if imaged with the ngVLA at 15~GHz.  The angular resolution is $\theta_{\mathrm{FWHM}} = 0.014^{\prime \prime}$.  Adapted from \citet{nyland+18}.}
\label{fig:NGC404_sim}
\end{figure*}

\subsection{Imaging Simulations}
We present ngVLA and JVLA simulations of a more distant analog to the confined jet with an extent of $\sim$10~pc in the center of NGC\,404 in Figure~\ref{fig:NGC404_sim}.  
To produce the input model, we scaled the image pixel size such that it corresponds to an NGC\,404 analog located at 10~Mpc, or $\sim$3X further than the true distance to NGC\,404.  
For the ngVLA simulations, we used the ngVLA configuration with 300 antennas and maximum baselines of $B_{\mathrm{max}} \sim$ 513~km as shown in \citet{nyland+18}.  The JVLA simulations were performed in the A-configuration, which has $B_{\mathrm{max}} = 36.4$~km.  

Figure~\ref{fig:NGC404_sim} illustrates the inability of the JVLA to spatially resolve the morphology of a 10~pc jet at a distance of 10~Mpc in the most extended A configuration.  This is in contrast to the simulated ngVLA observations, which successfully resolve the extended structure of a $\sim$10-pc-scale jet at a distance of 10~Mpc at 15~GHz.  As discussed in detail in \citet{nyland+18}, both the JVLA in the A configuration and the ngVLA would be able to detect emission from an NGC\,404 analog at a distance of 10~Mpc in a reasonable amount of on-source integration time ($\lesssim 15$~hr).  However, only the ngVLA would be able to spatially resolve the emission.  Given the low-luminosity of NGC\,404 ($L_{15\,{\rm GHz}} \sim 3.4 \times 10^{17}$~W~Hz$^{-1}$), spatially resolving the morphology of the source is crucial to distinguish its origin from nuclear star formation \citep{nyland+17}.  

\section{Summary}
In addition to placing constraints on the basic physical properties (e.g., extent, luminosity, and energy) of jetted AGNs in nearby low-mass galaxies, observations of the molecular ISM as traced by the CO(1--0) line at 115~GHz will probe the amount of energy transferred to the ISM and, in conjunction with supporting multiwavelength ancillary data from telescopes such as ALMA and the {\it James Webb Space Telescope}, its impact on the ambient star formation efficiency.  Thus, the unprecedented capabilities of the ngVLA will enable new advancements in our understanding of SMBH formation and the impact of jet-ISM feedback driven by SMBH seed analogs in the local Universe.

\bibliography{ngVLA_dwarf_galaxies_AGNs_v1}  

\end{document}